# Act of CVT and EVT In The Formation of Number-Theoretic Fractals

Pal Choudhury Pabitra[1], Sahoo Sudhakar [1], Nayak Birendra Kumar[2], and Hassan Sk. Sarif [2]

[1,] Applied Statistics Unit, Indian Statistical Institute, Kolkata, 700108, INDIA
[2.] P.G. Department of Mathematics, Utkal University, Bhubaneswar-751004

Email: pabitrapalchoudhury@gmail.com, sudhakar.sahoo@gmail.com, bknatuu@yahoo.co.uk & sarimif@gmail.com

**Abstract.** In this paper we have defined two functions that have been used to construct different fractals having fractal dimensions between 1 and 2. More precisely, we can say that one of our defined functions produce the fractals whose fractal dimension lies in [1.58, 2) and rest function produce the fractals whose fractal dimension lies in (1, 1.58]. Also we tried to calculate the amount of increment of fractal dimension in accordance with base of the number systems. And in switching of fractals from one base to another, the increment of fractal dimension is constant, which is 1.58, it's quite surprising!

## 1. Introduction:

In this paper we have tried to make an association between natural number and fractals, with the help of two defined transformations. Numbers corresponding to different number systems with base b (b=2, 3, 4…etc) signifies different fractals. Here, we have defined two new transformations called as "Carry Value Transformation (CVT) and Extreme Value Transformation (EVT)". With these two mapping we have generated fractals whose dimension lying in between the open interval (1, 2) [figure 7, 14]. It should be noted that we are traversing this interval discretely, but very densely also. That is for any given number from (1, 2) we could be able to give a fractal whose dimension is nearer to that given number. In our journey we have got a fractal whose dimension is 1.68, fortunately this very fractal dimension is the fractal dimension of music (Sri Lankan, Chariots of Fire) [7]. So this fractal could be a frame of lyrics for music. And undoubtedly there are a lot of fractals, which are of fractal dimension 1.68, and possibly these fractal-frames make new lyrics. In general, the fractal dimension of music is around the number 1.65, and our generated fractals could interpolate the number 1.65, as fractal dimension. In this paper also we have tried to calculate the amount of increment of fractal dimension in accordance with base of the number systems. And in switching of fractals from one base to another, the increment of fractal dimension is constant, which is 1.58, fractal dimension of Sierpinski Gasket.

   The organization of the paper is as follows. Section 2 discusses some of the basic concepts on fractals, fractal dimension, which are used in the subsequent sections. The concept of CVT is defined in section 3. In section 4, we have explored the formation of fractals in different bases of the number systems. And then we have generalized the concept of formation of fractals in any arbitrary bases of the number system. In section 5, we have explained the amount of increment of fractal dimension in accordance with base of the number systems.
In the next sections we have defined another transformation to generate fractals having fractal dimension lying in between (1, 1.58]. Also we have discussed formation of fractals in different bases of the number system and ultimately we have made it-generalized concept in any base of the number system as like for CVT we have done it. On highlighting other possible applications of CVT and some future research directions a conclusion is drawn in section 8.

## 2. Reviews of fundamentals of fractals

 For quite a long time the scientific community was very much worried due to our inability to describe the shape of cloud, a mountain, a coastline or a tree on using the traditional Euclidean Geometry. In nature, clouds are not really spherical, mountains are not conical, coastlines are not circular, even the lightning doesn't travel in a straight line. More generally, we would be able to conclude that many patterns of nature are so irregular and fragmented, that, compared with *Euclid Geometry* –a term, can be used in this regard to denote all of the standard geometry. Mathematicians have over the years disdained this challenge and have increasingly chosen to flee nature by devising theories unrelated to natural objects we can see or feel.

After a long time, responding to this challenge, Benoit Mandelbrot developed a new geometry of nature and implemented its use in a number of diverse arenas of science such as Astronomy, Biology, Mathematics, Physics, and Geography and so on [1, 2,3,4]. This new-born geometry can describe many of the irregular and fragmented (chaotic) patterns around us, and leads to full-fledged theories, by identifying a family of shapes, now-a-days which we people call 'FRACTALS'.

2.1 *Fractal dimension*

The fractal dimension alone does not give an idea of what "fractals" are really about Mandelbrot founded his insights in the idea of self similarity, requiring that a true fractal "fracture" or break apart into smaller pieces that resemble the whole. This is a special case of the idea that there should be a dynamical system underlying the geometry of the set. This is partly why the idea of fractals have become so popular throughout science; it is a fundamental aim of science to seek to understand the underlying dynamical properties of any natural phenomena. With this view, self similar fractals has been playing a dominant role, however it has now become more clearer that apart from self similarity,. Dynamical systems can produce many intricate shapes and behavior that occur throughout nature. There are different methods to calculate the fractal dimension of an object. But here we are only concentrating on Similarity dimension. Now let us try to define what fractal dimension (Similarity dimension) is. Given a self-similar structure [figure 3], there is a relation between the reduction factor (scaling factor) 'S' and the number of pieces 'N' into which the structure can be divided; and that relation is as follows…

$$N = 1/S^D, \text{ equivalently,}$$
$$\text{i.e. } D = \log(N)/\log(1/S)$$

This 'D' is called the Fractal dimension (Self-similarity dimension).

## 3. Carry Value Transformation (CVT)

3.1 *Formal definition of CVT*

The carry or overflow bits are usually generated at the time of addition between two n-bit strings. In the usual addition process, carry value is always a single bit and if generated then it is added column wise with other bits and not saved in its own place. But the carry value defined here are the usual carries generated bit wise and stored in their respective places as shown in fig 1.

$$\begin{array}{rl} \textit{carry value} = & c_n \quad c_{n-1} \ldots\ldots\ldots\ldots\ldots\ldots c_1 \quad 0 \\ a = & a_n \quad a_{n-1} \ldots\ldots\ldots\ldots\ldots a_1 \\ b = & b_n \quad b_{n-1} \ldots\ldots\ldots\ldots\ldots b_1 \\ a \oplus b = & a_n \oplus b_n \quad a_{n-1} \oplus b_{n-1} \ldots\ldots\ldots a_1 \oplus b_1 \end{array}$$

[Figure 1: Carry generated in $i^{th}$ column is saved in $(i-1)^{th}$ column]

Thus to find out the carry value we perform the bit wise XOR operation of the operands to get a string of sum-bits (ignoring the carry-in) and simultaneously the bit wise ANDing of the operands to get a string of carry-bits, the latter string is padded with a '0' on the right to signify that there is no carry-in to the LSB. Thus the corresponding decimal value of the string of carry bits is always an even integer.

Now we can give a precise definition of CVT as follows:

Let $B = \{0,1\}$ and CVT is a mapping defined as $CVT : (B_n \times B_n) \to B_{n+1}$ where $B_n$ is the set of strings of length $n$ on $B = \{0,1\}$. More specifically, if $a = (a_n, a_{n-1},...,a_1)$ and $b = (b_n, b_{n-1},...,b_1)$ then $CVT(a,b) = (a_n \wedge b_n, a_{n-1} \wedge b_{n-1},...,a_1 \wedge b_1, 0)$ is an (n+1) bit string, belonging to $\mathbb{Z}$, set of non-negative integers, and can be computed bit wise by logical AND operation followed by a 0, which denotes no carry is generated in the LSB at the time of addition procedure. In other words, CVT is a mapping from $\mathbb{Z} \times \mathbb{Z} \rightarrow \mathbb{Z}$, where $\mathbb{Z}$ is set of non-negative integers [7].

Example:

Suppose, we want to get the CVT of the numbers $(13)_{10} \equiv (1101)_2$ and $(14)_{10} \equiv (1110)_2$. Both are 4-bit numbers. The carry value is computed as follows:

```
Carry:  1 1 0 0 0
Augend:   1 1 0 1
Addend:   1 1 1 0
XOR:      0 0 1 1
```

[Figure 2: Carry generated in $i^{th}$ column is saved in $(i-1)^{th}$ column]

Conceptually, in the general addition process the carry or overflow bit from each stage (if any) goes to the next stage so that, in each stage after the first (i.e. the LSB position with no carry-in), actually a 3-bit addition is performed instead of a 2-bit addition by means of the full adder. Instead of going for this traditional method, what we do is that we perform the bit wise XOR operation of the operands (ignoring the carry-in of each stage from the previous stage) and simultaneously the bit wise ANDing of the operands to get a string of carry-bits, the latter string is padded with a '0' on the right to signify that there is no carry-in to the LSB (the overflow bit of this ANDing being always '0' is simply ignored). In our example, bit wise XOR gives $(0011)_2 \equiv (3)_{10}$ and bit wise ANDing followed by zero-padding gives $(11000)_2 \equiv (24)_{10}$. Thus $CVT(1101,1110) = 11000$ and equivalently in decimal notation one can write $CVT(13,14) = 24$. In the next section we have used the carry value in decimal to construct the CV table.

**4. Generation of fractal using CVT in different bases of number system**

A table is constructed that contains only the carry values (or even terms) defined above between all possible integers a's and b's arranged in an ascending order of x and y-axis respectively. We observe some interesting patterns in the table. We would like to make it clear how the CV-table is constructed.

Step 1. Arrange all the integers 0 1 2 3 4 5 6 ... (as long as we want) in ascending order and place it in both, uppermost row and leftmost column in a table.

Step 2. Compute $CVT(a,b)$ as mentioned in 3.1 and store it in decimal form in the (a, b) position.

Then we look on the pattern of any integer, and we have made it color. This shows a very beautiful consistent picture, which we see as a fractal as by shown below.

4.1 *Generation of fractals*

Let us do find the fractals in different domain of number system with the help of CVT.

4.1.1 *Production of fractal in binary (2-nary) number system*

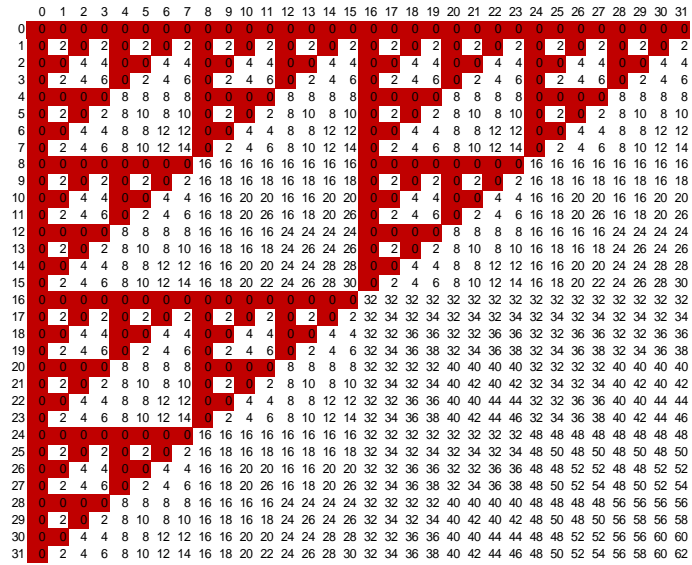

[Figure 3: a fractal structure on using CVT of different integer values]

*Dimension of this fractal*
For this fractal [see figure 3], number of self-similar copies N=3 and scaling factor S=1/2, where l is the initial length. So, referring the discussion in 2.1, Fractal dimension D is given by
$$3=1/(1/2)^D$$
Or D= log3/log2 ≈ 1.585

This is same as the dimension of Sierpinski triangle. Thus CVT fractal as obtained by us can be regarded as a relative to *Sierpinski* triangle [6].

4.1.2 *CV- table in Ternary (3-nary) number system*
Here we are applying CVT on the domain of ternary (3-nary) number system and we are having the following table. It is mentioned that the procedure can be verbatim copied just by replacing binary by ternary (3-nary), as we have discussed earlier to construct CVT table in binary number system. Let us demonstrate how we are computing CVT of two numbers in ternary number system.

Suppose, we want to get CVT of the numbers $(13)_{10} \equiv (111)_3$ and $(14)_{10} \equiv (112)_3$. Both are 3-digit numbers. The carry value is computed as follows:

$$
\begin{array}{rl}
\text{Carry:} & 0\ 0\ 1\ 0 \\
\text{Augend:} & 1\ 1\ 1 \\
\text{Addend:} & 1\ 1\ 2 \\
\text{Addition process in ternary:} & 2\ 2\ 0
\end{array}
$$

Therefore, CVT(13,14)= $(0010)_3$ =3.

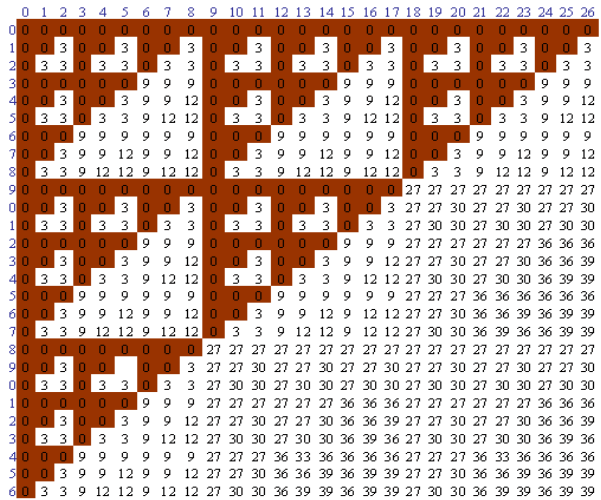

[Figure 4: a fractal structure on using CVT of different integer values in ternary number system]

*Dimension of this fractal*
For this fractal [see figure 4], number of self-similar copies N=3 and scaling factor S=1/2, where l is the initial length. So, referring the discussion in 2.1, Fractal dimension D is given by
$$6 = 1/(1/3)^D$$
Or D= log6/log3 $\approx$ 1.630929

4.1.3 *CV- table in 4-nary number system*
Here we are applying CVT on the domain of 4-nary number system and we are having the following table. It is mentioned that the procedure can be verbatim copied just by replacing binary by 4-nary, as we have discussed earlier to construct CVT table in binary number system.

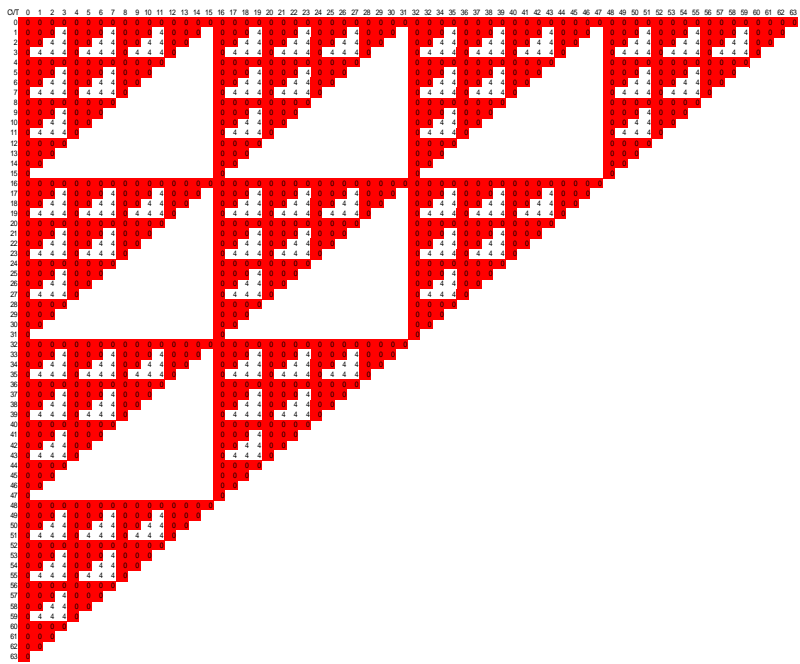

[Figure 5: a fractal structure on using CVT of different integer values in quadruple number system]

*Dimension of this pattern*
For this fractal [see figure 5], number of self-similar copies N=3 and scaling factor S=1/2, where l is the initial length. So, referring the discussion in 2.1, Fractal dimension D is given by
$$10 = 1/(1/4)^D$$
Or D= log10/log4=1.6609

4.1.4 *CV- table in 5-nary number system*
Here we are applying CVT on the domain of ternary (3-nary) number system and we are having the following table. It is mentioned that the procedure can be verbatim copied just by replacing binary by ternary (3-nary), as we have discussed earlier to construct CVT table in binary number system.

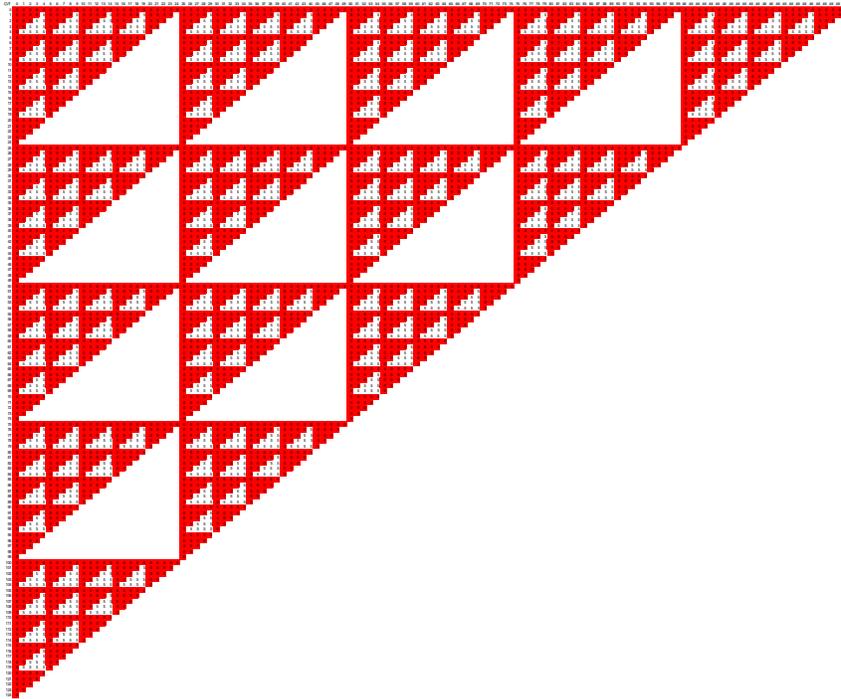

[Figure 6: a fractal structure on using CVT of different integer values in 5-nary number system]

*Dimension of this pattern*
For this fractal [see figure6], number of self-similar copies N=3 and scaling factor S=1/2, where l is the initial length. So, referring the discussion in 2.1, Fractal dimension D is given by
$$15 = 1/(1/5)^D$$
Or D= log15/log5=1.682606

It should be noted that fractal dimension of music (Sri Lankan, Chariots of Fire) is also 1.68[7]. So this fractal could be a frame of lyrics for music. In general fractal dimension of arbitrary music is nearly 1.65. And undoubtedly there are a lot of fractals which are of fractal dimension 1.68, and possibly these fractal-frames make new lyrics.

Similarly it could be possible to generate fractals in any base of the number system.

4.2 *Generalization of the concept in arbitrary number system*
Now, we are warmed up to construct the fractal in different number system just following the above procedure. It is also cleared that, when we have calculated the similarity dimension of those fractal in each case we have observed the scaling factor is 1/n and self-similar copies is (1+2+3+…n), for n-nary number system. So let us define our self-similarity dimension formula as follows…

Here, Number of self − similar copies,

$$N = \sum_{j=1}^{n} j, \ \& \ Scaling, S = \frac{1}{n}, for\ n-nary\ number\ system$$

$$Fractal\ dimension, S_D(n) = \left\{ \frac{\log(\sum_{j=1}^{n} j)}{\log n} \right\}$$

$$i.e. S_D(n) = \left\{ \log\left(\frac{n(n+1)}{2}\right) \Big/ \log n \right\}$$

*Theorem 1*: The fractal dimension $S_D$ converges to the topological dimension (Euclidian dimension) 2 as the base 'n' of the number system diverges to infinity.

Proof: *Let us try to find the limit when n tends to infinity,*

$$\lim_n S_D(n) = \lim_n \left\{ \log\left(\frac{n(n+1)}{2}\right) \Big/ \log n \right\} \ \left(\frac{\infty}{\infty}\ form\right)$$

$$= \lim \left\{ \frac{\left[\frac{2(2n+1)}{2n(n+1)}\right]}{\frac{1}{n}} \right\}; by\ L'hospital\ rule = \lim\left\{\frac{2n^2+n}{n^2+n}\right\} = 2$$

So, starting from the binary number system the fractal dimension of the generated fractal will go on increasing with the increase of the base of the number system and finally it converges to the topological dimension 2.

Let us consider a list of such dimensions…

**Table 1**.A table with fractal dimension of fractals according as base of the number system

| Base of the number system | Fractal dimension of the obtained fractal |
|---:|---:|
| 2 | 1.584962501 |
| 3 | 1.630929754 |
| 4 | 1.660964047 |
| 5 | 1.682606194 |
| 6 | 1.699180325 |
| 7 | 1.712414374 |
| 8 | 1.723308334 |
| 9 | 1.73248676 |
| 10 | 1.740362689 |
| 11 | 1.747221736 |
| 12 | 1.75326861 |
| 13 | 1.758654413 |
| 14 | 1.763493463 |
| 15 | 1.767874074 |
| 16 | 1.77186571 |
| 17 | 1.77552387 |
| 18 | 1.778893508 |
| 19 | 1.782011483 |
| 20 | 1.784908344 |
| 21 | 1.787609657 |
| 22 | 1.790137008 |
| 23 | 1.792508765 |
| 24 | 1.794740674 |
| 25 | 1.796846321 |
| 26 | 1.798837498 |
| 27 | 1.800724501 |
| 28 | 1.802516365 |
| 29 | 1.804221054 |

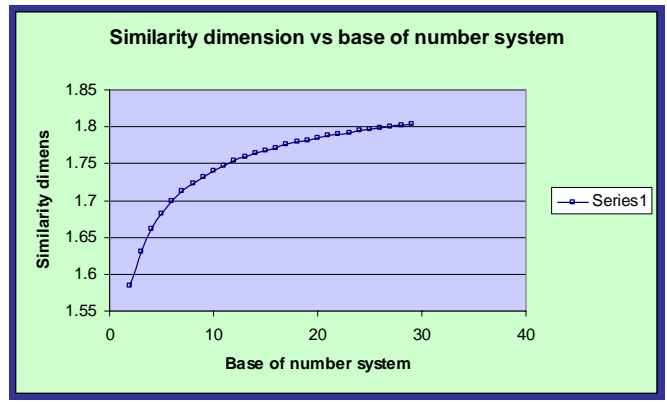

[Figure 7: Graph shows how fractal dimension increases in accordance with base of the number system, using table 1]

**5. Increment of fractal dimension of CVT fractals**

Let us try to obtain the increment of fractal dimension of the CVT fractals in switching from one base to another base of the number system. In binary number system we have the fractal of fractal dimension 1.58, and in ternary number system we have the fractal of fractal dimension 1.63. So algebraically, it seems the increment of fractal dimension is (1.63-1.58) = 0.05. But it is really not!
To obtain the increment we proceed as follows…
First of all we paste one (n-1)-nary CVT fractal to n-nary CVT fractal. Next, the overflowed portion is extracted which can be seen as a self-similar figure. These self-similar pieces derived from overflowed portion would lead to another fractal with some fractal dimension. What we observe, this very measure (fractal dimension) is the actual increment of fractal dimension in switching from one base to another base of the number system.

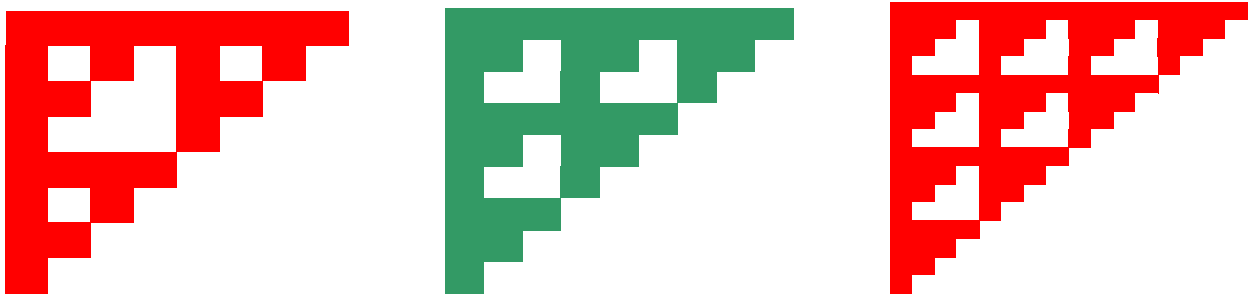

[8(a) Binary CVT fractal generator] [8(b) Ternary CVT fractal generator] [8(c) 4-nary CVT fractal generator]

5.1. *Binary generator over Ternary generator*
Here we paste binary CVT fractal over ternary CVT fractal [8(a),8(b)]. Next, the overflowed portion can be seen in figure 9, which is a self-similar figure as the CVT fractals were self-similar. These self-similar pieces derived from overflowed portion would lead to another fractal with fractal dimension 1.58 as shown below.

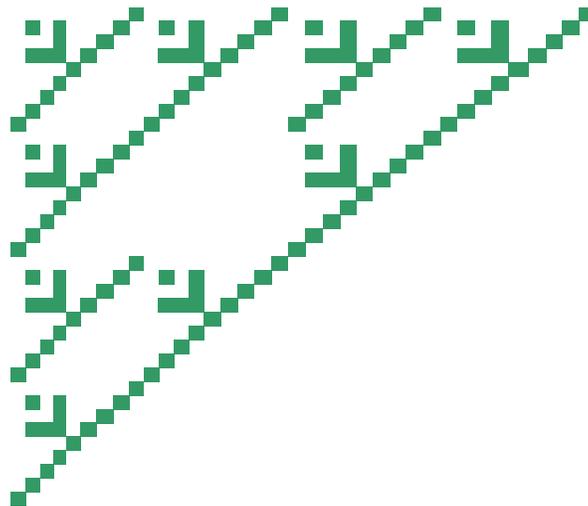

[Figure 9.Binary CVT generator over Ternary CVT generator lead to another generator for a fractal.]

This generator leads to another fractal; let us calculate the fractal dimension of the generated fractal as follows…

Dimension of this fractal
For this fractal, N=3, S=1/2, where l is the initial length.

Fractal dimension D is given by …     $3=1/(1/2)^D$
Or D= log3/log2 ≈ 1.585

This is same as the dimension of Sierpinski triangle. Thus CVT fractal as obtained by us can be regarded as a relative to *Sierpinski* triangle.
Therefore we could say that the amount of increment is likely 1.58.

5.2. *Ternary generator over 4-nary generator*
Here we paste ternary CVT fractal over 4-nary CVT fractal [8(b),8(c)]. Next, the overflowed portion can be seen in figure 10, which is a self-similar figure as the CVT fractals were self-similar. These self-similar pieces derived from overflowed portion would lead to another fractal with fractal dimension 1.58 as shown below.

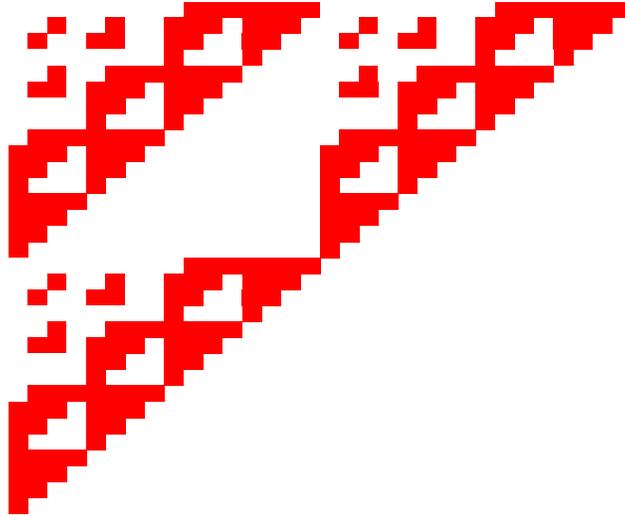
[Figure 10.Ternary CVT generator over 4-nary CVT generator lead to another generator for a fractal.]

This generator also leads to another fractal; let us calculate the fractal dimension of the generated fractal as follows…

*Dimension of this fractal*
For this fractal, N=3, S=1/2, where l is the initial length.

Fractal dimension D is given by …     $3=1/(1/2)^D$
Or D= log3/log2 ≈ 1.585

This is same as the dimension of Sierpinski triangle. Thus CVT fractal as obtained by us can be regarded as a relative to *Sierpinski* triangle. Here also we are having the increment same as above.

In fact, in general we could make a conjecture that if we like to paste an n-nary CVT generator over an (n+1)-nary CVT generator, then we will be able to have another generator which led to the fractal of fractal dimension 1.58, i.e. the attractor fractal is Sierpinski Gasket.
So, the amount of increment is constant, which is 1.58.

So far we have discussed we are in position to achieve the fractals having fractal dimension lying in between [1.58, 2), with the help of CVT. But let us try to define another map named as Extreme Value Transformation (EVT) to achieve fractals whose fractal dimension lying in between (1, 1.58].

**6. Extreme Value Transformation (EVT)**

6.1 *Formal definition of EVT*
Now we can give a precise definition of EVT as follows:
Let $B = \{0,1\}$ then EVT is a mapping defined as $E_{max}VT: B_n \times B_n \to B_n$ where $B_n$ is the set of strings of length $n$ on $B = \{0,1\}$. More specifically, if $a = (a_n, a_{n-1},...,a_1)$ and $b = (b_n, b_{n-1},...,b_1)$ then
$E_{max}VT: \mathbb{Z} \times \mathbb{Z} \to \mathbb{Z}$ where a, b being non-negative integers.

$$E_{max}(a,b) = ((\max(a_n, b_n), (\max(a_{n-1}, b_{n-1})... \max(a_1, b_1)))$$

is an n bit string and can be computed bit wise maximum.

Now, people may have in confusion that why we have concentrate on bit-wise maximum in defining EVT. But we like to make you clear that if we use the bit wise minimum then also we will be having same fractal if we ignore the orientation (each of them what we are getting $E_{ma.x}VT$). That is we are producing a pair of functions to produce the same fractals.

Example:
Suppose, we want the EVT of the numbers $(13)_{10} \equiv (1101)_2$ and $(14)_{10} \equiv (1110)_2$. Both are 4-bit numbers. The extreme value is computed as follows…

$$13 = 1101$$
$$14 = 1110$$
$$EVT\ (13, 14) = 1110$$

## 7. Generation of fractal using EVT in different bases of number system

A table is constructed that contains only integer defined above between all possible integers a's and b's arranged in an ascending order of x and y-axis respectively. We observe some interesting patterns in the table. We would like to make it clear how the EV-table is constructed.

Step-1. Arrange all the integers 0 1 2 3 4 5 6 ... (as long as we want) in ascending order and place it in both, uppermost row and leftmost column in a table.
Step-2. Compute EVT (a, b) as mentioned in 6.1 and store it in decimal form in the (a, b) position.
Then we look on the pattern of any integer, and we have made it color. This shows a very beautiful consistent picture, which we see as a fractal as by shown below.

*7.1 Generation of fractal*
Let us do find the fractals in different domain of number system with the help of CVT.

*7.1.1 Production of fractal in binary (2-nary) number system*

|    | 0 | 1 | 2 | 3 | 4 | 5 | 6 | 7 | 8 | 9 | 10 | 11 | 12 | 13 | 14 | 15 |
|----|---|---|---|---|---|---|---|---|---|---|----|----|----|----|----|----|
| 0  | 0 | 1 | 2 | 3 | 4 | 5 | 6 | 7 | 8 | 9 | 10 | 11 | 12 | 13 | 14 | 15 |
| 1  | 1 | 1 | 3 | 3 | 5 | 5 | 7 | 7 | 9 | 9 | 11 | 11 | 13 | 13 | 15 | 15 |
| 2  | 2 | 3 | 2 | 3 | 6 | 7 | 6 | 7 | 10| 11| 10 | 11 | 14 | 15 | 14 | 15 |
| 3  | 3 | 3 | 3 | 3 | 7 | 7 | 7 | 7 | 11| 11| 11 | 11 | 15 | 15 | 15 | 15 |
| 4  | 4 | 5 | 6 | 7 | 4 | 5 | 6 | 7 | 12| 13| 14 | 15 | 12 | 13 | 14 | 15 |
| 5  | 5 | 5 | 7 | 7 | 5 | 5 | 7 | 7 | 13| 13| 15 | 15 | 13 | 13 | 15 | 15 |
| 6  | 6 | 7 | 6 | 7 | 6 | 7 | 6 | 7 | 14| 15| 14 | 15 | 14 | 15 | 14 | 15 |
| 7  | 7 | 7 | 7 | 7 | 7 | 7 | 7 | 7 | 15| 15| 15 | 15 | 15 | 15 | 15 | 15 |
| 8  | 8 | 9 | 10| 11| 12| 13| 14| 15| 8 | 9 | 10 | 11 | 12 | 13 | 14 | 15 |
| 9  | 9 | 9 | 11| 11| 13| 13| 15| 15| 9 | 9 | 11 | 11 | 13 | 13 | 15 | 15 |
| 10 | 10| 11| 10| 11| 14| 15| 14| 15| 10| 11| 10 | 11 | 14 | 15 | 14 | 15 |
| 11 | 11| 11| 11| 11| 15| 15| 15| 15| 11| 11| 11 | 11 | 15 | 15 | 15 | 15 |
| 12 | 12| 13| 14| 15| 12| 13| 14| 15| 12| 13| 14 | 15 | 12 | 13 | 14 | 15 |
| 13 | 13| 13| 15| 15| 13| 13| 15| 15| 13| 13| 15 | 15 | 13 | 13 | 15 | 15 |
| 14 | 14| 15| 14| 15| 14| 15| 14| 15| 14| 15| 14 | 15 | 14 | 15 | 14 | 15 |
| 15 | 15| 15| 15| 15| 15| 15| 15| 15| 15| 15| 15 | 15 | 15 | 15 | 15 | 15 |

[Figure 11: a fractal structure on using CVT of different integer values]

*Dimension of this fractal*
For this fractal [Figure 11], N=3, S=1/2, where l is the initial length.

Fractal dimension D is given by ...    $3 = 1/(1/2)^D$
$$\text{Or } D = \log 3/\log 2 \approx 1.585$$
This is same as the dimension of Sierpinski triangle. Thus CVT fractal as obtained by us can be regarded as a relative to *Sierpinski* triangle [4].

7.1.2 *CV- table in Ternary (3-nary) number system*
Here we are applying EVT on the domain of ternary (3-nary) number system and we are having the following table. It is mentioned that the procedure can be verbatim copied just by replacing binary by ternary (3-nary), as we have discussed earlier to construct EVT table in binary number system.

|    | 0  | 1  | 2  | 3  | 4  | 5  | 6  | 7  | 8  | 9  | 10 | 11 | 12 | 13 | 14 | 15 | 16 | 17 |
|----|----|----|----|----|----|----|----|----|----|----|----|----|----|----|----|----|----|----|
| 0  | 0  | 1  | 2  | 3  | 4  | 5  | 6  | 7  | 8  | 9  | 10 | 11 | 12 | 13 | 14 | 15 | 16 | 17 |
| 1  | 1  | 1  | 2  | 4  | 4  | 5  | 7  | 7  | 8  | 10 | 10 | 11 | 13 | 13 | 14 | 16 | 16 | 17 |
| 2  | 2  | 2  | 2  | 5  | 5  | 5  | 8  | 8  | 8  | 11 | 11 | 11 | 14 | 14 | 14 | 17 | 17 | 17 |
| 3  | 3  | 4  | 5  | 3  | 4  | 5  | 6  | 7  | 8  | 12 | 13 | 14 | 12 | 13 | 14 | 15 | 16 | 17 |
| 4  | 4  | 4  | 5  | 4  | 4  | 5  | 7  | 7  | 8  | 13 | 13 | 14 | 13 | 13 | 14 | 16 | 16 | 17 |
| 5  | 5  | 5  | 5  | 5  | 5  | 5  | 8  | 8  | 8  | 14 | 14 | 14 | 14 | 14 | 14 | 17 | 17 | 17 |
| 6  | 6  | 7  | 8  | 6  | 7  | 8  | 6  | 7  | 8  | 15 | 16 | 17 | 15 | 16 | 17 | 15 | 16 | 17 |
| 7  | 7  | 7  | 8  | 7  | 7  | 8  | 7  | 7  | 8  | 16 | 16 | 17 | 16 | 16 | 17 | 16 | 16 | 17 |
| 8  | 8  | 8  | 8  | 8  | 8  | 8  | 8  | 8  | 8  | 17 | 17 | 17 | 17 | 17 | 17 | 17 | 17 | 17 |
| 9  | 9  | 10 | 11 | 12 | 13 | 14 | 15 | 16 | 17 | 9  | 10 | 11 | 12 | 13 | 14 | 15 | 16 | 17 |
| 10 | 10 | 10 | 11 | 13 | 13 | 14 | 16 | 16 | 17 | 10 | 10 | 11 | 13 | 13 | 14 | 16 | 16 | 17 |
| 11 | 11 | 11 | 11 | 14 | 14 | 14 | 17 | 17 | 17 | 11 | 11 | 11 | 14 | 14 | 14 | 17 | 17 | 17 |
| 12 | 12 | 13 | 14 | 12 | 13 | 14 | 15 | 16 | 17 | 12 | 13 | 14 | 12 | 13 | 14 | 15 | 16 | 17 |
| 13 | 13 | 13 | 14 | 13 | 13 | 14 | 16 | 16 | 17 | 13 | 13 | 14 | 13 | 13 | 14 | 16 | 16 | 17 |
| 14 | 14 | 14 | 14 | 14 | 14 | 14 | 17 | 17 | 17 | 14 | 14 | 14 | 14 | 14 | 14 | 17 | 17 | 17 |
| 15 | 15 | 16 | 17 | 15 | 16 | 17 | 15 | 16 | 17 | 15 | 16 | 17 | 15 | 16 | 17 | 15 | 16 | 17 |
| 16 | 16 | 16 | 17 | 16 | 16 | 17 | 16 | 16 | 17 | 16 | 16 | 17 | 16 | 16 | 17 | 16 | 16 | 17 |
| 17 | 17 | 17 | 17 | 17 | 17 | 17 | 17 | 17 | 17 | 17 | 17 | 17 | 17 | 17 | 17 | 17 | 17 | 17 |

[Figure 12: a fractal structure on using CVT of different integer values in ternary number system]

*Dimension of this fractal*
For this fractal [Figure 12], N=5, S=1/3, where l is the initial length.

Fractal dimension $S_D$ is given by ...
$$5 = 1/(1/3)^D$$
$$\text{Or } S_D = \log 5/\log 3 \approx 1.46497$$

7.1.3 *EV- table in 4-nary number system*
Here we are applying EVT on the domain of 4-nary number system and we are having the following table. It is mentioned that the procedure can be verbatim copied just by replacing binary by 4-nary, as we have discussed earlier to construct EVT table in binary number system.

[Figure 13: a fractal structure on using CVT of different integer values in 4-nary number system]

*Dimension of this pattern*
For this periodic pattern [Figure 13], N=7, S=1/4, where 1 is the initial length.

Fractal dimension D is given by ... $7 = 1/(1/4)^D$
Or D= log7/log4=1.403677

*7.2 Generalization of the concept in arbitrary number system*
Now, we are warmed up to construct the fractal in different number system just following the above procedure. It is also cleared that, when we have calculated the similarity dimension of those fractal in each case we have observed the scaling factor is 1/n and self-similar copies is (1+2+3+...n), for n-nary number system. So let us define our self-similarity dimension formula as follows...

$$\text{Here, Number of self} - \text{similar copies,}$$
$$N = n + (n-1) = 2n - 1$$
$$S = \frac{1}{n}, \text{for } n - \text{nary number system}$$

$$\text{Fractal dimension}, S_D(n) = \left\{\frac{\log(2n-1)}{\log n}\right\}$$

*Theorem 2:* The fractal dimension $S_D$ converges to the topological dimension (Euclidian dimension) 1 as the base 'n' of the number system, diverges to infinity.

Proof:
Let us try to find the limit when n tends to infinity,
$$\lim_n S_D(n) = \lim_n \left\{\frac{\log(2n-1)}{\log n}\right\} \left(\frac{\infty}{\infty} \text{ form}\right)$$

$$= \lim \left\{ \left[\frac{2}{(2n-1)}\right] \Big/ \frac{1}{n} \right\} \text{ by L'Hospital rule}$$

$$= \lim \left\{ \frac{2n}{2n-1} \right\} = 1$$

i. e. $\lim_{n \to \infty} S_D(n) = 1$

So, starting from the binary number system the fractal dimension of the generated fractal will go on decreasing with the increase of the base of the number system and finally it converges to the topological dimension 1.
Let us consider a list of such dimensions…

**Table 2**. A table with fractal dimension of fractals according as base of the number system

| Base of the number system | Fractal dimension of the obtained fractal |
|---|---|
| 2 | 1.584962501 |
| 3 | 1.464973521 |
| 4 | 1.403677461 |
| 5 | 1.365212389 |
| 6 | 1.338290833 |
| 7 | 1.318123223 |
| 8 | 1.302296865 |
| 9 | 1.289450962 |
| 10 | 1.278753601 |
| 11 | 1.269664473 |
| 12 | 1.261815697 |
| 13 | 1.254947126 |
| 14 | 1.248868992 |
| 15 | 1.24343922 |
| 16 | 1.238549078 |
| 17 | 1.234113756 |
| 18 | 1.230066012 |
| 19 | 1.226351756 |
| 20 | 1.222926921 |
| 21 | 1.219755197 |
| 22 | 1.21680636 |
| 23 | 1.214055019 |
| 24 | 1.211479669 |
| 25 | 1.209061955 |
| 26 | 1.206786106 |
| 27 | 1.20463848 |
| 28 | 1.202607215 |
| 29 | 1.195425616 |

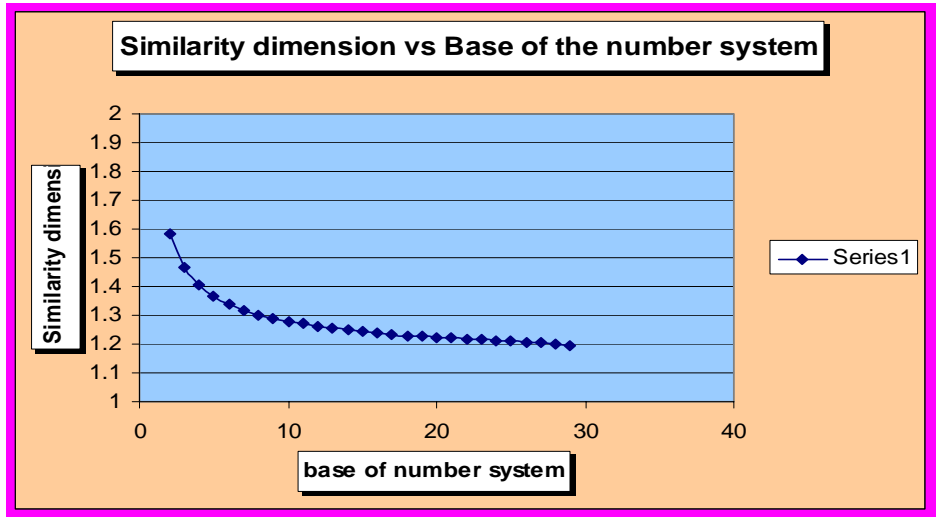

[Figure 14: Graph shows how fractal dimension increases in accordance with base of the number system]

Here it is noted that similar analysis for decrement of fractal dimension in switching of EVT fractals from one base to another base of the number system could be drawn.

**8. Conclusion and Future Research directions**

So far we have observed how CVT and $E_{Extreem}VT$ produce fractals. In this regard one natural question may be raised as to whether some other functions would be able to produce the same fractals. We are happy to produce those functions also. Already we have observed that in extreme value transformation we have considered the bit wise maximum, instead of bit-wise maximum if we consider bit wise minimum we will be having those fractals but different orientation. Then we are in firm conviction that corresponding to CVT there is also another twin like function, which also produces those fractals although we have not yet analyzed them. Our current effort will be to analyze those twin functions algebraically and thus finding the broader spectrum of these fractals.

**References.**

[1] Mandelbrot, B.B. 1982. *The fractal geometry of nature*. New York
[2] P. P. Choudhury, S. Sahoo, and M. Chakraborty, 2008, Implementation of Basic Arithmetic Operations Using Cellular Automaton, ICIT-08, IEEE Computer Society Press, proceedings of 11[th] International Conference on Information Technology, Bhubaneswar, India, pp 79-80, Dec. 2008.
[3] Pietgen, H. O., Jurgeens, H., Saupe, D., 1992. *Chaos and Fractals New Frontiers of Science*, ISBN 3-540-97903-4, Springer Verlag.
[4] Ghosh, A.K., Choudhury, P. P., Choudhury, R., 2003 Production of fractals by various means and measuring their dimensions with probable explanation, Laser Horizon, Journal of Laser Science and Technology Centre (LASTEC), vol. **6/No. 2**, pp.50-59.
[5] Bogomolny, A., Cut The Knot! An interactive column using Java applets http://www.cut-the-knot.org/ctk/Sierpinski.shtml
[6] Meyer Perrin S. Columbia University, New York, Fractal dimension of music.
[7] P. P. Choudhury, S. Sahoo, B. K Nayak, and Sk. S. Hassan, 2009, Carry Value Transformation: It's Application in Fractal Formation, *IEEE International Advanced Computing Conference (IACC 2009*), Patiala, India, **6-7 March**, pp 2613-2618, 2009.